\documentclass[pra,twocolumn,amsmath,amssymb,groupedaddress,superscriptaddress]{revtex4-1}
\usepackage{graphics}
\usepackage{amsmath}
\usepackage[dvips]{graphicx}
\usepackage{float,graphicx}
\usepackage{epsfig}
\usepackage{epstopdf}
\usepackage{amssymb}
\usepackage{epsfig}
\usepackage{graphicx}
\usepackage{amsmath}
\usepackage{amsfonts}
\usepackage{mathrsfs}
\newcommand{\ket}[1]{\left\vert #1 \right\rangle}
\newcommand{\bra}[1]{\left\langle #1 \right\vert}

\def\>{\rangle}
\def\<{\langle}
\def\ket#1{|#1\>}
\def\bra#1{\<#1|}

\def\e{\rm e}
\def\be{\begin{equation}}

\def\ee{\end{equation}}

\def\be{\begin{equation}}
\def\ee{\end{equation}}
\def\bea{\begin{eqnarray}}
\def\eea{\end{eqnarray}}
\usepackage{bm}
\usepackage{graphicx}

\begin{document}

\title{Time-frequency resolved ultrafast spectroscopy techniques using wavelet analysis}

\author{Javier Prior}
\affiliation{Departamento de F\'isica Aplicada, Universidad Polit\'ecnica de Cartagena, Cartagena 30202, Spain}
\author{Enrique Castro}
\affiliation{Departamento de F\'isica Aplicada, Universidad Polit\'ecnica de Cartagena, Cartagena 30202, Spain}
\author{Alex W. Chin}
\affiliation{Theory of Condensed Matter Group, University of Cambridge,
J J Thomson Avenue, Cambridge, CB3 0HE, United Kingdom}
\author{Javier Almeida}
\affiliation{Institut f\"{u}r Theoretische Physik, Albert-Einstein-Allee 11, Universit\"{a}t Ulm, D-89069 Ulm, Germany}
\author{Susana F. Huelga}
\affiliation{Institut f\"{u}r Theoretische Physik, Albert-Einstein-Allee 11, Universit\"{a}t Ulm, D-89069 Ulm, Germany}
\author{Martin B. Plenio}
\affiliation{Institut f\"{u}r Theoretische Physik, Albert-Einstein-Allee 11, Universit\"{a}t Ulm, D-89069 Ulm, Germany}

\begin{abstract}

New experimental techniques based on non-linear ultrafast spectroscopies have been developed over the last few years, and have been demonstrated  to provide powerful probes of quantum dynamics in different types of molecular aggregates, including both natural and artificial light harvesting complexes. Fourier transform-based spectroscopies have been particularly successful, yet 'complete' spectral information normally necessitates the loss of all information on the temporal sequence of events in a signal. This information though is particularly important in transient or multi-stage processes, in which the spectral decomposition of the data evolves in time. By going through several examples of ultrafast quantum dynamics, we demonstrate that the use of wavelets provide an efficient and accurate way to simultaneously acquire \emph{both} temporal and frequency information about a signal, and argue that this greatly aids the elucidation and interpretation of physical process responsible for non-stationary spectroscopic features, such as those encountered in coherent excitonic energy transport.

\end{abstract}

\date{\today}
\maketitle

\section{Introduction}

Experimental data are usually obtained by perturbing a system and then measuring responsive degrees of freedom, such as optical, magnetic, phononic, density fluctuations, etc.  Such responses are normally recorded by the experimental apparatus in either the temporal or the frequency domain. Direct measurement in the frequency domain (spectral analysis) generates detailed information about the dynamics and characteristic excitations of the system under study, and provides invaluable input to theoretical analysis. Through the application of the Fourier transform (FT), such information can also be obtained from the measurement of signals in the time domain but then key information about the temporal order of the 'events' in the response is lost. On the other hand, the form of the experimental signal in the temporal domain often appears very complex and/or noisy, making it hard to pick out important features captured by the spectral analysis.  This loss of information is highly undesirable if the dynamical processes being investigated are transient and non-stationary, that is, they have a spectral decomposition that \emph{evolves} in time - especially if these changes are sudden. In such a case, which is typical for interacting systems, the FT is effectively an average of these evolving spectra, which may greatly obscure the information that was present in the apparently erratic original time-trace signal.

In recognition of the fact that many, if not most important physical process are non-stationary, the technique of the Wavelet Transform (WT) has emerged as an important tool to study
transient processes \cite{VanderBerg04}, particularly in the context of mechanical vibrations, acoustics, digital signal processing and data compression. The key to its high impact across these large research fields is its highly optimized time-frequency
decomposition of a signal, which allows the temporal evolution of a system's frequency spectrum to be followed with uncertainties close to the Heisenberg limit posed by the energy-time uncertainty relation \cite{Mallat99}.

Part of our motivation for presenting this technique, whose use in physics and physical chemistry is currently less developed than in the disciplines above, is the recent development and application of ultrafast non-linear spectroscopy techniques to probe coherent dynamics in molecules, solids and liquids. An excellent example is Two Dimensional Photon Echo Spectroscopy (2D PES) \cite{Hybl98,Brixner05}, a relatively recent technique that is able to decongest the often highly overlapping optical excitation spectra of molecular complexes, and which reveals
inter- and intra-molecular dynamics on femto-to-picosecond time scales \cite{darius,fleming}. Moreover, it is also able, in principle, to separate out electronic
and vibrational signals \cite{greg}, which is extremely difficult to do with linear absorption
techniques. As a topical example of where these techniques have lead to major re-thinking of how photophysical processes occur, we mention the recently observed and long-lasting (ps) coherent dynamics in some biological pigment-protein complexes (PPCs) found in light harvesting antennae \cite{NP13,Chin12,Engel07,Tiwari13,olaya}. Here, the extensive,  but partially obscured, information available through 2D PES is key to answering the fascinating question of how quantum coherence is supported in biological environments and how it might actually help drive photosynthetic light reactions in the forward direction. The well-established technique of wavelet analysis may provide a new strategy to extract more of this crucial information.

In the next section we will present briefly an intuitive introduction to the wavelet transformation, while an extended exposition of some technical details is deferred to \cite{VanderBerg04,Mallat99,wt}. The subsequent sections of this paper provide a range of illustrative examples of its analytical utility.  In section II and III we will apply the wavelet analysis to address two problems of relevance to ultrafast optical or vibrational spectroscopy. Namely, the vibronic coupling in a dissipative excitonic dimer and the determination of the lineshape function in linear absorption spectra.

In both cases, the main advantage that we shall demonstrate is the ability of the WT to effectively spread out the complete information contained in either the frequency or temporal signal, e.g. in non-linear spectroscopies, to give a representation that allows this full information to be interpreted in terms of physical processes. Indeed, as we shall show, the WT allows even simple, linear (one dimensional) measurements to be spread into detailed multidimensional representations, possibly allowing important physics to be obtained in cheap, simple, but effective set ups. Such effectively one-dimensional spectroscopies have recently provided unprecedented information about \emph{single} molecule dynamics in photosynthetic complexes \cite{Hulst2013}, and WT applied to such experiments could yet provide further insight.

The WT is not the only way to carry out a time-frequency analysis. Another option is using the Short-Time Fourier Transform (STFT) \cite{Kreisbeck2013}, which is the Fourier transform over short periods of time of a signal. The STFT uses the same time windows to calculate the frequencies in a given time, so the absolute resolution in frequency is the same for all the frequencies. In contrast, the WT realizes a multi resolution analysis \cite{Mallat89} by using different time intervals (scales) for different frequencies. For low frequencies the WT uses long timescales which yields high frequency resolution while for high frequencies the WT uses short scales thus achieving better temporal localization. Therefore, the WT is optimally adapted to each of the frequencies involved in the signal \cite{Mallat99}. As it will be apparent later on, it is this flexibility that makes WT ideally tailored to analyze transient, non stationary dynamics.

\section{Brief Introduction to the Wavelet transformation}

\begin{figure}
\includegraphics[scale=0.38]{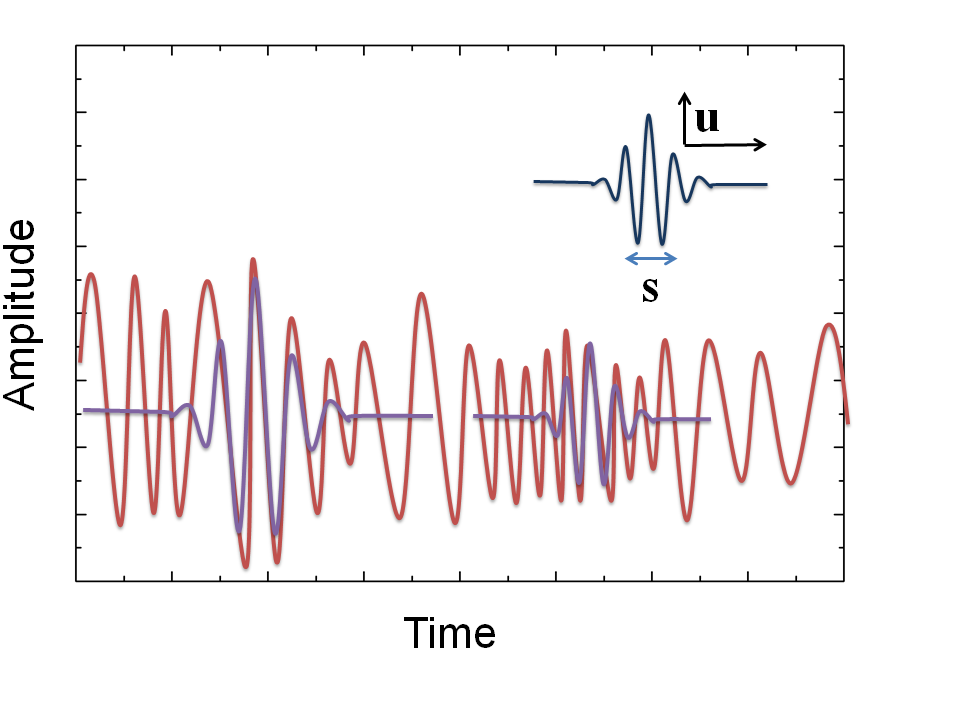}
\caption{Typical wavelet representation. We illustrate pictorially two possible expansions of a given signal $f(t)$ (red curve) in terms of its inner product with a wavelet atom
located at two different positions, as determined by the $u$ coordinate, and for two different values of the scale parameter $s$.}
\label{Fig0}
\end{figure}

The WT refers to several transformations and signal processing techniques \cite{wt,Harrop02} that have in common that they use a zero mean and
short-time oscillating function $\psi$, called a 'mother' wavelet, to decompose a one- or  multi-dimensional real signal into different frequency bands, known in the wavelet literature as scales.
This mother wavelet function is translated by $u$ and dilated by the scale $s$, giving the wavelet 'atom' function, which provides the effective basis for the transformation:

\begin{equation}
\psi_{u,s}(t)=\frac{1}{\sqrt{s}}\psi\left(\frac{t-u}{s}\right).
\end{equation}

The two most common transformations are the so called Discrete Wavelet Transform (DWT) and the Continuous
Wavelet Transform (CWT) \cite{VanderBerg04}. The DWT decompose the signal into several
frequency bands and is much used for data and image compression and de-noising signals. The CWT, which is the transform we shall use
in this paper, is based on an expansion of a temporal signal $f(t)$ via the inner product of the function with a wavelet atom \cite{Mallat99}:

\begin{eqnarray}
\label{CWT}
CWT_{f}(u,s)&=&\int^{+\infty}_{-\infty}f(t)\frac{1}{\sqrt{s}}\psi^{*}\left(\frac{t-u}{s}\right)dt \nonumber \\&=&
\int^{+\infty}_{-\infty}f(t)\psi^{*}_{u,s}(t)dt \nonumber \\
&=&\left\langle f, \psi_{u,s}(t)\right\rangle
\end{eqnarray}

The translation parameter $u$ indicates where the wavelet atom is centered on
the temporal axis, while the scale parameter $s$ controls
how the relative width of the wavelet atom compared to the mother wavelet function; this effectively determines its duration (see Fig. \ref{Fig0}). This duration of the wavelet atom can be seen as its effective period,
so that the frequency of the wavelet atom will be proportional to $1/s$.

Usually, the CWT is made over the whole temporal axis, scanning $u$ over the full temporal domain of $f(t)$ for fixed scale $s$. The operation
is repeated, if necessary, with a different scale to obtain the CWT for the whole
temporal axis and a chosen interval of $s$ values.

Eq. \ref{CWT} shows that the CWT of a function $f(t)$ quantifies the overlap between $f$ and a wavelet atom at a particular time and of a particular frequency. As shown in Fig. \ref{Fig0}, this overlap (and the strength of the signal in the corresponding wavelet spectrum, known as a scalogram) would be higher if the typical frequencies of the signal at this time match those of the wavelet atom. By calculating the scalogram (see \cite{VanderBerg04,Mallat99,wt}), it is possible to obtain the instantaneous frequencies of $f$ by looking at which wavelet scales contribute the most to the wavelet decomposition at given time $t$.

\begin{figure}
\includegraphics[scale=0.47]{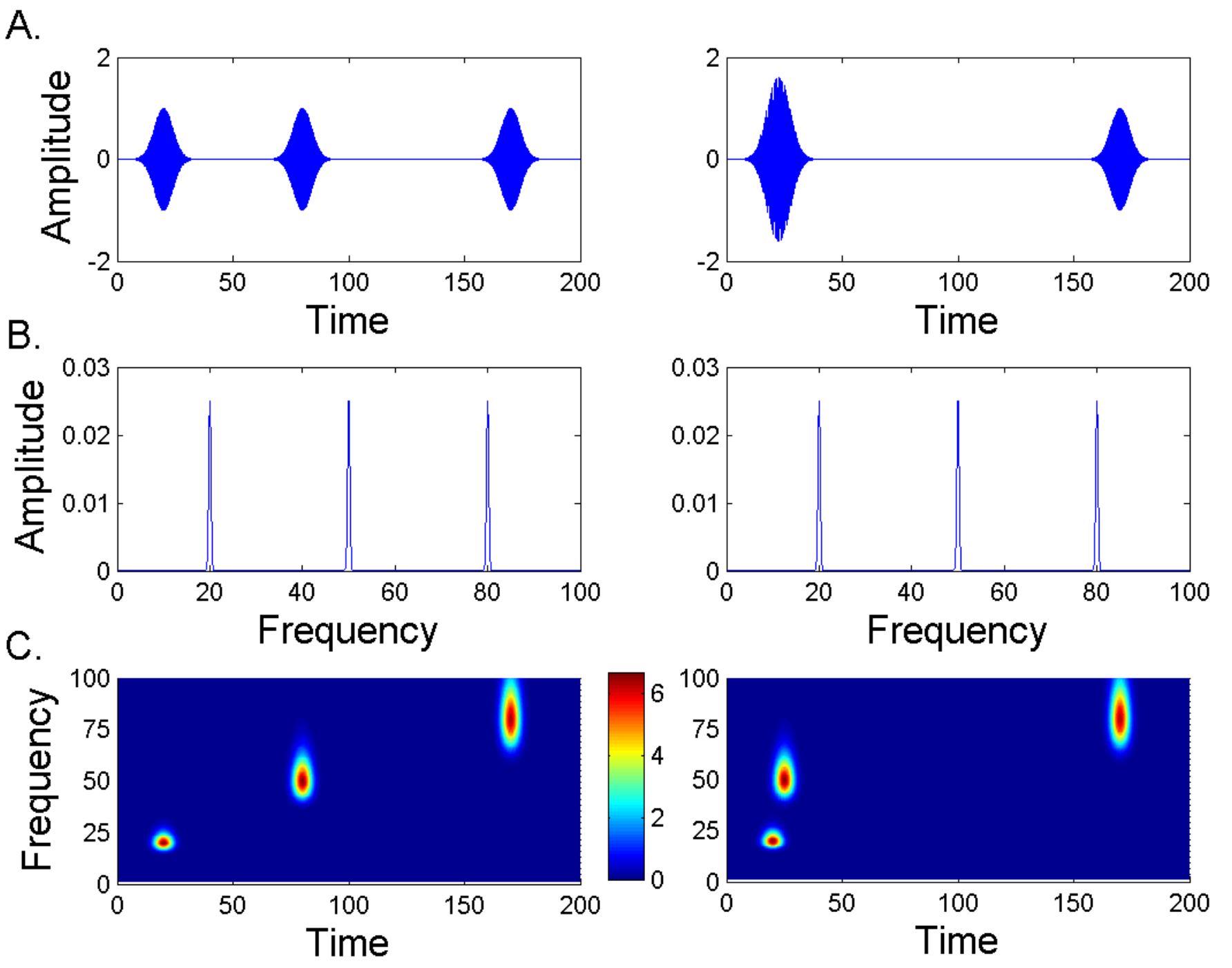}
\caption{Illustration of the WT approach. In the top plot 2.A, we represent two different signals described by  Eg. (\ref{eq:funcion})
and for the parameters written in the text. Amplitudes and times are displayed in arbitrary units. The FT of
these functions are identical, as illustrated in the middle panel 2.B. In the bottom plots 2.C we represent the scalogram of the two signals, showing that they become distinguishable when
frequency-time resolution, as provided by the wavelet transform, is accessible.}
\label{Fig1}
\end{figure}

An intuitive example to understand WT is illustrated in Fig. \ref{Fig1}.  Consider the signal function

\begin{equation}
f(t)=\sum_{i=1}^3\sin (\omega_i t)\exp^{-\frac{1}{2}(\frac{t-\mu_i}{\sigma_i})^2}
\label{eq:funcion}
\end{equation}
which is the sum of three sinusoidal functions with frequencies $\omega_i$ ($i=1-3$) which are temporally located at times $\mu_i$. The corresponding variances are $\sigma_i$.

We choose the frequencies of the signal to be $\omega_1=20$, $\omega_2=50$ and $\omega_3=80$. We now generate two different signals
by taking the Gaussian envelopes
centered at $\mu_1=20$, $\mu_2=80$ and $\mu_3=170$ (Fig. \ref{Fig1}.A left)
and $\mu_1=20$, $\mu_2=25$ and $\mu_3=170$ (Fig. \ref{Fig1}.A right), with $\sigma=4$).
With these choices, the $\mu_i$ values are such that there is not overlap between
the different sine functions (Fig. \ref{Fig1}.A left) , while the center values of the first and second
sinusoidals do overlap in the second case. As a result, the FT of the considered signals are identical, as shown in Fig. \ref{Fig1}.B
where we obtain the three expected peaks at $20$, $50$ and $80$ for both signals.
However, we can break down this equivalence when resolving the temporal component of the frequency domain.
In Fig. \ref{Fig1}.C we represent the WT analysis which allows the relevant signal frequencies to feature
at different times. Now we can clearly observe the non overlapping components in the first case, while the second signal exhibits a partial temporal overlap of two characteristic frequencies.

In order to explain the possible relevance of using wavelets in the context of ultrafast spectroscopy, we will now present some examples of non-stationary dynamics of relevance to the study of molecular aggregates, like for instance pigment-protein complexes (PPCs) involved in energy transfer which have been recently probed using multidimensional spectroscopy \cite{naturechem}. Note that our aim here is merely to show how the physics of various toy models of PPCs may be further understood using WT analysis; discussions of how these processes contribute to the light reactions or what role they suggest for the presence of coherent effects in biological systems is beyond our scope, though we hope that wavelet analysis may provide future insight into these interesting problems.

\section{A model structure: two level system coupled to a harmonic oscillator}

\begin{figure}
\includegraphics[scale=0.45]{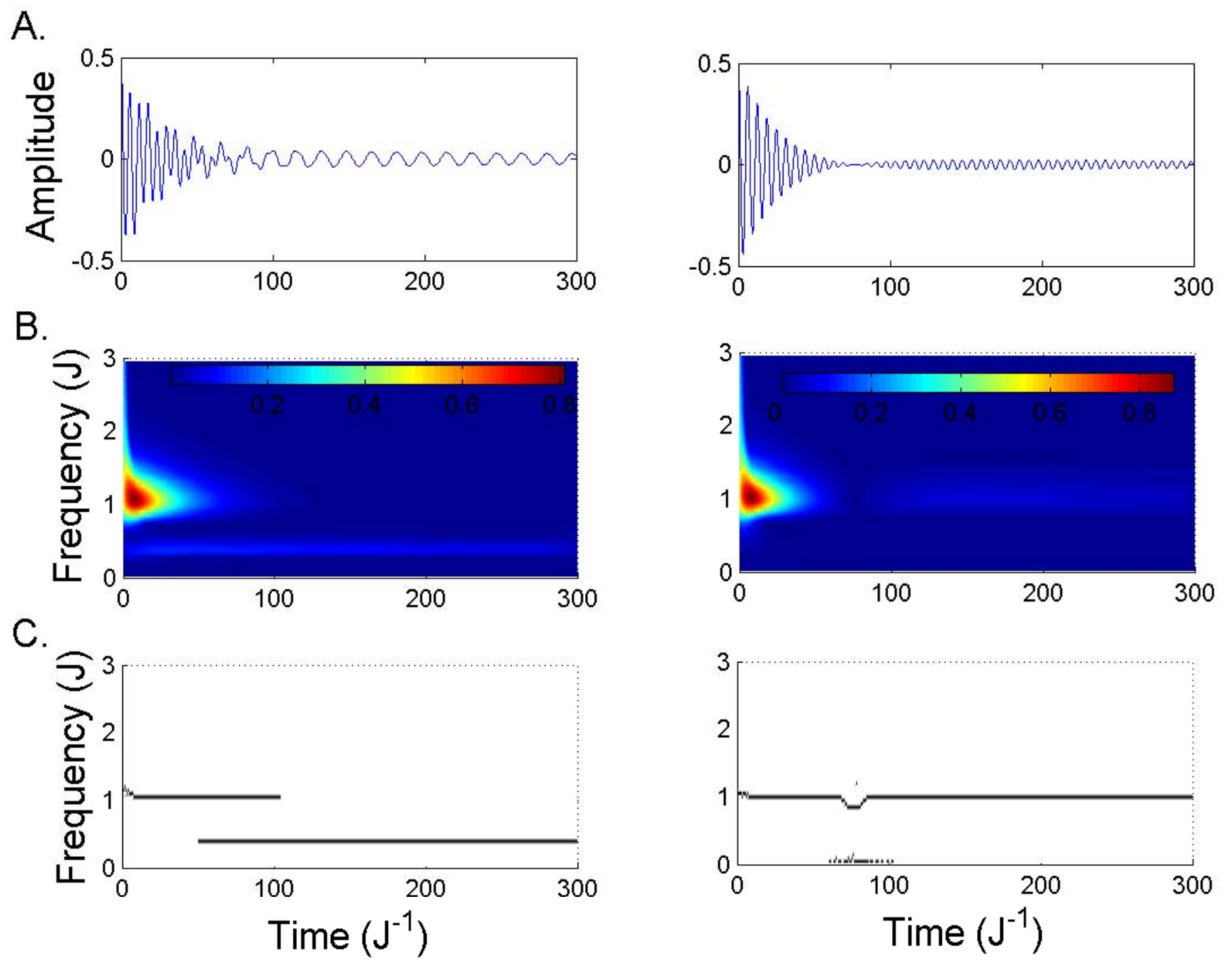}
\caption{\textbf{Electronic Coherence Dynamics}.
Considering the system dynamics as given by Eq.\ref{eq:Hvib_diag}, and for the initial state $\rho(0)=\ket{u}\bra{u}\otimes \ket{0}\bra{0}$, we analyze the electronic coherence dynamics
for the case of weak electron-phonon coupling (left A-C panels, where $g/J=0.015$) versus a stronger coupling regime (right A-C panels, $g/J=0.15$).
The top panels represent the amplitude of the coherence of the
  electronic degrees of freedom as quantified by $\rho_{e_{1}e_{2}}(t)=Re\left[ {\rm Tr}\{\rho(t)\ket{e_{1}}\bra{e_{2}}\}\right]$. In both cases the dephasing rate
  $\gamma$ has been fixed to $\gamma/J=0.01$.
The scalogram of both signals is represented in the central B. panels while the bottom C. plot shows the wavelet analysis of the largest frequencies
resulting from the scalogram analysis, we plot, at each time, the frequencies of those local maxima that are within 20$\%$ of the global maximum at that time.}
\label{fig:coherences}
\end{figure}

The wavelet analysis is
particularly well-suited to physical systems with multi-frequency dynamics. These
scales are typically associated with different degrees of freedom, and may also evolve due their interactions and interplay. Pigment-protein complexes (PPCs) of photosynthetic organisms, and more generally, vibronic molecular systems, provide a nice example to illustrate this behaviour. Here, the strong interaction of electronic (pigment) dynamics  -which carry the solar energy absorbed from sunlight - and the vibrational motions of the molecular and protein structure leads to an intrinsic and non-stationary multi-scale dynamics.  As a toy model to illustrate this, we describe a simple electronic-vibrational system consisting of
an electronically coupled dimer (ECD) \cite{NP13} whose excited delocalized states are coupled through a single quantum
harmonic oscillator of frequency $\omega$. Following textbook descriptions of linear electron-vibrational coupling, the effective Hamiltonian for a single optical excitation (Frenkel exciton) in the dimer in such a coupled system is given by (for $\hbar=1$) \cite{GronBook}:
\begin{equation}
H_{\rm X-V}=
J \hat{\sigma}^z +
g\hat{X}\hat{\sigma}^x + \omega a^{\dagger} a
\label{eq:Hvib_diag}
\end{equation}
where $a^{\dagger}$, $a$ are bosonic creation and annihilation operators, respectively, describing the vibrational mode in second quantisation. The displacement operator of the vibrational mode $\hat{X}$ is defined as
$\hat{X}=a^{\dagger} + a$ and $g$ is the coupling strength between the vibration and excitonic dimer. The Pauli matrices above describe the two possible single-exciton excited states of the ECD defined in the
'delocalized' basis, i.e,  $\sigma^z\equiv
\ket{e_{1}}\bra{e_{1}}-\ket{e_{2}}\bra{e_{2}}$ and $\sigma^x\equiv
\ket{e_{1}}\bra{e_{2}}+\ket{e_{2}}\bra{e_{1}}$, where $|e_{i}\rangle$ denotes a state where the dimer has a single excitation in its $i$th excited state. The parameter $J$ is the coherent splitting of the delocalised excited states. More generally, our dimer model has the same structure as the Rabi model, itself a single-vibration variant of the celebrated spin-boson model, both of which also describe numerous other physical systems in both quantum optics and the condensed phase \cite{weissBook, wallsBook}. The origin of the states involved in our excitonic representation of Eq. (\ref{eq:Hvib_diag}) are shown in Fig.\ref{fig:coherences}, though we note that considerable approximations have been made in the reduction of the real system to the toy model we shall analyze for the sake of showing how wavelets could be applied to such dynamics. The role of the interplay between coherent excitonic and vibrational coupling in the dissipative environment of real PPCs is a much more complex problem, which includes the issue of how coherence is maintained on ps timescales in these systems \cite{Prior10, NP13, Chin12}. Nevertheless, and in anticipation of later applications to more sophisticated treatments of the dynamics, we now present how wavelet analysis of the excitonic coherence in our toy model elucidates the basic physics.
In order to study the dynamics
of the electronic coherences between the states $|e_{1}\rangle$ and $|e_{2}\rangle$, we will make use of a Lindblad
master equation \cite{Rivas11} for the density matrix of our exciton-vibrational system with a pure dephasing term directly coupled to the excitonic
wavefunction (i.e. diagonal in the excitonic basis). That is:
\begin{multline}
\frac{d\rho (t)}{dt}=-\frac{i}{\hbar}
[H_{X-V},\rho(t)]+ \gamma \mathscr{L}_{\rm deph}(\rho(t))
,\\
\text{ with }
\mathscr{L}_{\rm deph}(\rho)\equiv
\hat{\sigma}^z\rho\hat{\sigma}^z-\rho,
\label{eq:ME_coherences}
\end{multline}
where the constant $\gamma$ is the dephasing rate of excitonic coherence. The initial state of the following simulations is $\rho(0)=|u\rangle\langle u|\otimes|0\rangle\langle 0|$, where $\ket{u}=1/\sqrt{2}(\ket{\e_{1}} - \ket{e_{2}})$ and $|0\rangle$ is the vacuum state of the vibrational mode. Such a state corresponds to a anti-symmetric superposition of the exciton states with maximum initial excitonic coherence, and is uncorrelated with the vibrational state. In all simulations we measure energies in units of J, setting $J=1$ throughout.

In Fig.~\ref{fig:coherences} we have plotted the evolution of the
amplitude of the excitonic coherence, defined as $\rho_{e_{1}e_{2}}(t)=Re\left[ {\rm Tr}\{\rho(t)\ket{e_{1}}\bra{e_{2}}\}\right]$. To understand the dynamics, we can exploit a semiclassical analogy for the relatively weak exciton-vibration coupling considered here \cite{NP13}. The initially excited coherence will oscillate with its characteristic frequency ($2J$, the energy difference of excitonic excited states) but is damped by the pure dephasing. Such oscillatory coherence, in our model, transiently drives the (undamped) vibrational mode causing it to oscillate at its natural frequency (and with a relative phase to the excitonic coherence oscillations). However, as the initial excitonic coherence dies away due to damping, the back-action of the long-lasting oscillations of the vibrational mode - which is itself driving the excitonic coherence at the \emph{mode's} vibrational frequency - becomes dominant in the exciton coherence dynamics. As the dynamics is linear, we therefore expect the exciton coherence dynamics to contain two frequency components over time \cite{NP13,Chin12}, an initial excitonic oscillation at the excitonic energy splitting ($2J$) and a later "mode-driven" response close to the natural frequency of the vibrational mode.

\begin{figure}
\includegraphics[scale=0.45]{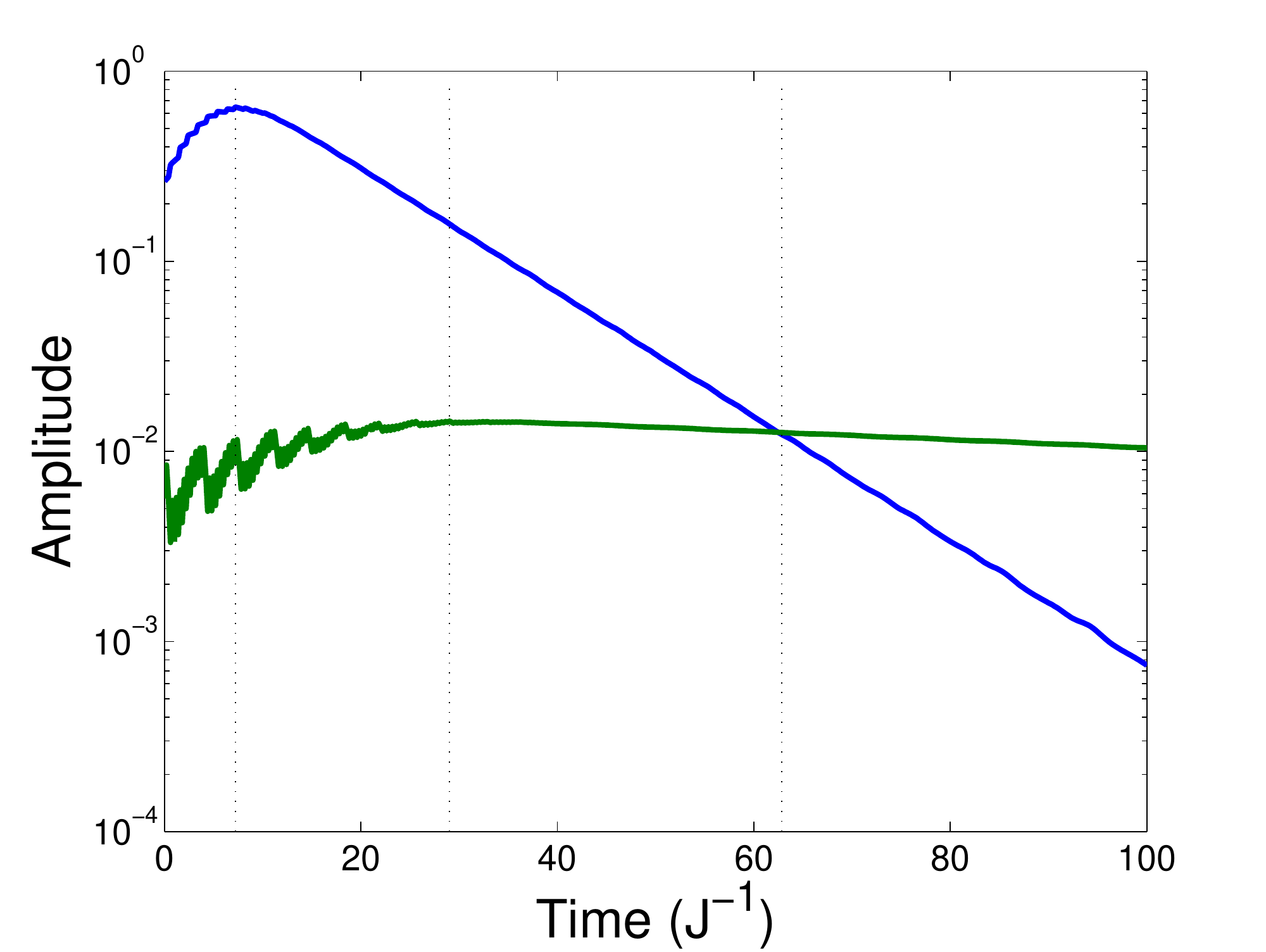}
\caption{Initial time-evolution of the high frequency wavelet amplitude at $\omega=1.05J$ (blue line) and the low frequency wavelet amplitude at $\omega=0.41J$ (green line) shown in the left panel of Fig. \ref{fig:coherences}.C. Vertical dotted lines highlight the times at which these peaks reach their maximum amplitude, and the region where the low-frequency response becomes dominant in the dynamics is also marked. }
\label{FigFreq}
\end{figure}

Fig.~\ref{fig:coherences}.A shows results for an off-resonant coupling to the selected excitonic transiton, with a vibrational frequency of $\omega=0.4J$ and $g=0.015J$. The time-trace of the coherence clearly shows the qualitative switch-over between the initial and mode-driven coherence oscillations (as evidenced by the change of frequency), but the wavelet analysis reveals even more detail. As it can be seen when looking at panels 3.B and 3.C, where we represent the scalogram of the signal and the resulting dominant frequency WT, wavelets allow us to look at the fates of the different frequency components independently. The initial high-frequency (excitonic) component dies away after about $t=100J^{-1}$ but its amplitude becomes comparable to the mode-driven component of the dynamics after about $t=40J^{-1}$ (Fig. \ref{fig:coherences}.C left), which also shows that both responses do co-exist. This mode-driven component then extends over a much longer time period due to the lack of direct dephasing of the vibrational mode. In Fig. (\ref{FigFreq}), we exploit our access to time-resolved spectral information to zoom in on the early time evolution of the high and low frequency wavelet bands. We clearly observe the time-delayed appearance of the low -frequency mode-driven response (relative to the peak of the initial exciton response at high frequency), revealing that this response was itself created by the initial excitonic oscillations acting on the mode. This must be the case for the initial conditions we have used here, but such an observation of a time-lagged response might be very useful in disentangling signals emerging from dynamical exciton-vibration interactions and situations where the vibrational modes are directly excited in experiments and would thus be driving an excitonic response from $t=0$.

In addition to revealing the dynamical cross-over from damped initial coherence oscillations to mode-driven, long-lasting coherence, the wavelet technique is also sensitive to the relative phase of the responses. This is illustrated in the right side of Fig.~\ref{fig:coherences}, which shows results for a vibrational mode that is resonant ($\omega=2J$) with the exciton energy difference. The mode-driven response at long times and the initial excitonic oscillations now have the same frequency, and as these - as discussed above - overlap in time, they may \emph{interfere}, according to their relative phase. The time trace of Fig.~\ref{fig:coherences}.A shows a distinct interval of suppressed oscillation ($t\approx75$), after which the oscillations grow again to a constant amplitude in the mode-driven regime.

The appearance of this suppressed region can again be understood intuitively by a simple semiclassical argument. The excitons initially drive the modes at resonance, thus we expect its amplitude response to be $\approx \pi/2$ out of phase with the coherence oscillations. These vibrational oscillations then drive the exciton coherence at its resonant frequency, and therefore we would expect the mode-driven component of the exciton coherence to be $\approx\pi$ out of phase with the initial oscillations. Such oscillations will destructively interfere, leading to the region of suppressed oscillation where the dominant contribution to the signal switches from the initial to mode-driven coherences. Moreover, this destructive interference mechanism will lead to a more rapid transfer of the initial coherence into the longer-lasting mode-driven form, as can be seen by comparing the lifetimes of the high frequency features in the wavelet plots of a and b. Alternatively, this behaviour could be described in terms of the initial excitation of a superposition of new normal modes formed by the resonant mixing of exciton and vibration states, this also leads to interference in the dynamics, and - in principle - a splitting of the dynamical frequencies. However, the small coupling and damping used in this example prevent direct resolution of the new mode frequencies.

\begin{figure}
\includegraphics[scale=0.30]{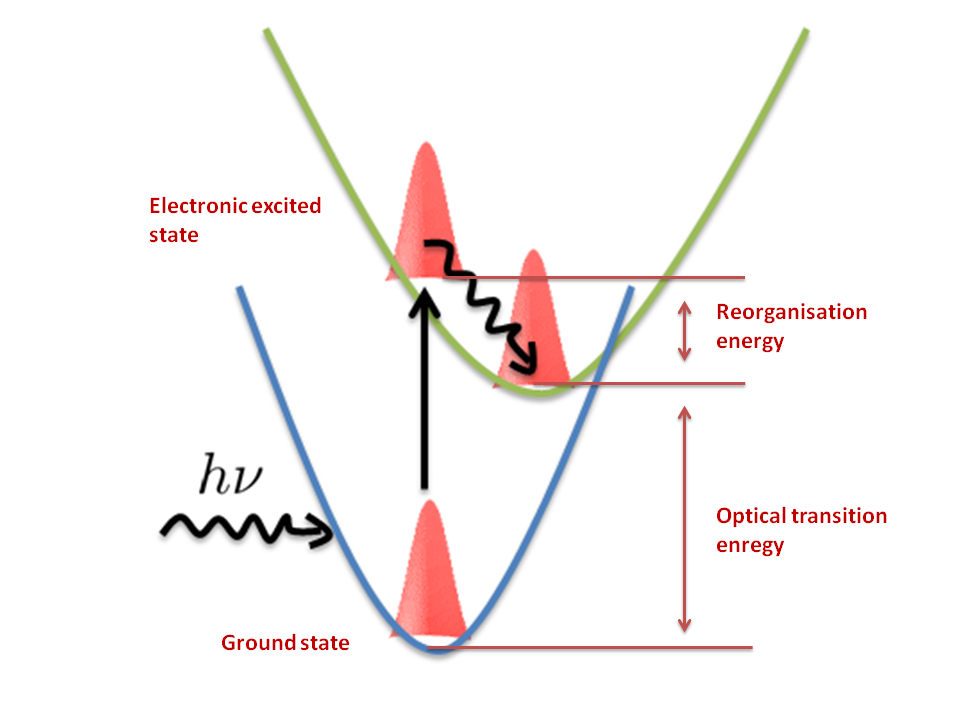}
\caption{Optical (vertical) excitation of the electronic ground state results in an excited state (assumed to be a linear displacement of the ground state potential energy surface) that is not an equilibrium molecular geometry. As the molecular co-ordinates relax (a generalised, a collective co-ordinate is shown for illustration, the energy gap between ground and excited states changes, resulting in a changing spectroscopic response. This change in absorption frequency is known as the time-dependent Stokes shift.}
\label{FigSF}
\end{figure}

The model system analyzed in this section illustrates the use of wavelet analysis to decongest linear dynamics containing multiple contributions which can appear at different frequencies, times and with different phases. However, each contributions had a characteristic and unchanging frequency, which we now contrast, in our final example, which presents wavelet analysis of an optical system with a single \emph{time-evolving} frequency.

\section{Environmental relaxation in linear absorption spectra }

We will now examine the effect of a time-dependent Stokes shift of an single (monomer), optically excited molecular pigment (generically known as a chromophore) on its linear absorption spectrum. Fig.~\ref{FigSF} shows the physical meaning of this spectral feature, which is normally modeled, much like the previous example, as arising from a linear coupling to a macroscopic bath of harmonic oscillators. The total Hamiltonian can be written as:

\begin{figure}
\includegraphics[scale=0.45]{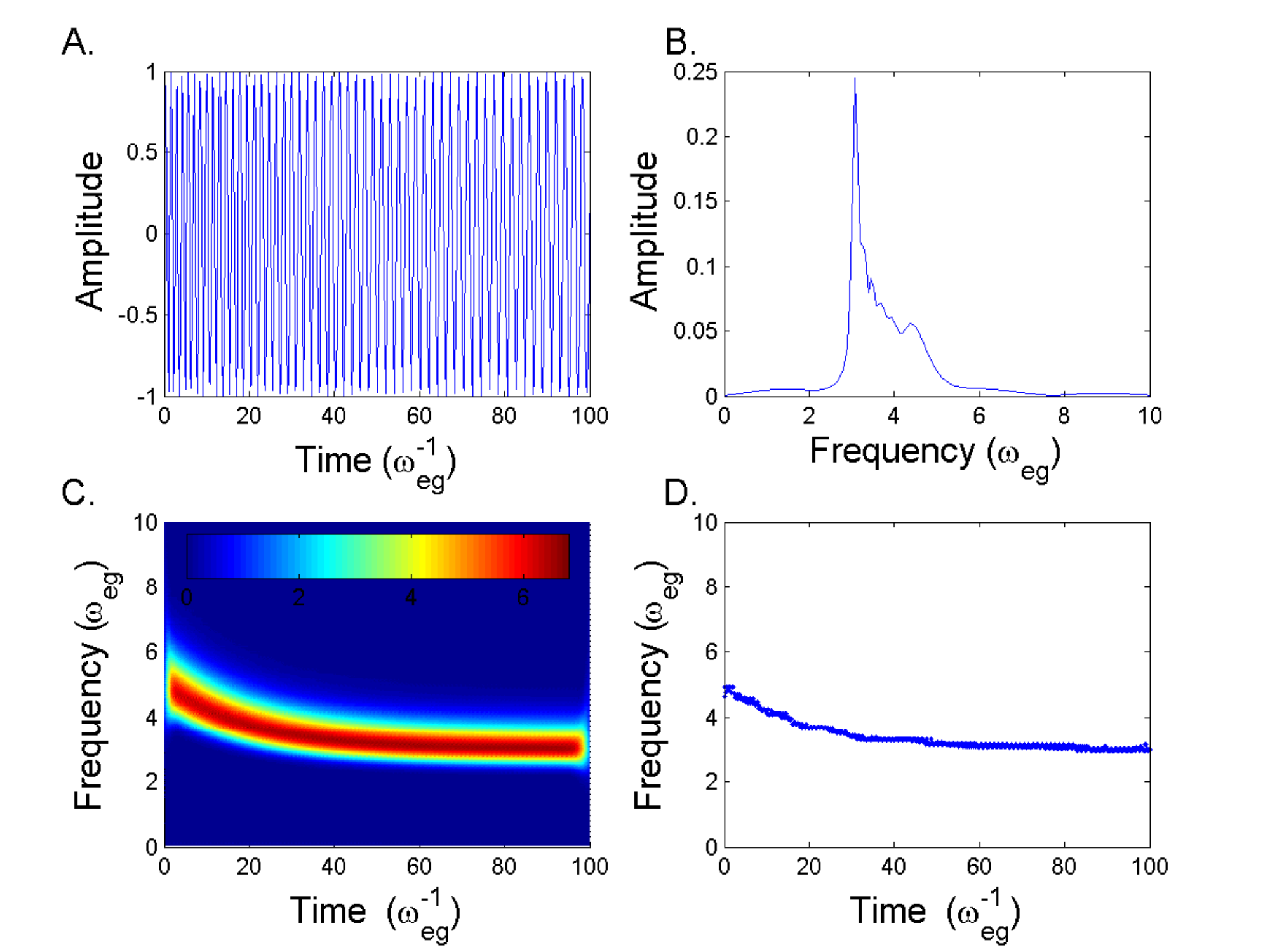}
\caption{Line Shape function with the parameter $G_{Re}=0$ no damping function. This is an artificial case of the line shape with zero amplitude decay. In the Fourier analysis we can see that the main frequency component is at the relaxed ground-excited energy gap frequency  $\Omega_{ab}$. The other parameter values of the line shape function are $S=2$, $\Omega_{eg}=3$ and $\Omega_D=0.05$ }
\label{W0}
\end{figure}

\begin{equation}
H=\frac{\omega_{eg}}{2}\sigma_z+\sum_\xi \Omega_\xi a_\xi^+a_\xi^-+\sum_\xi g_\xi\sigma^z(a_\xi^+ + a_\xi^-)
\label{Hamilt}
\end{equation}
and we define the spectral density of the environment \cite{weissBook} as

\begin{equation}
J(\omega)=\sum_{xi} g_\xi^2\delta(\omega-\omega_\xi).
\end{equation}

The absorption spectrum $D_{abs}(\omega)$ is computed from the Fourier transform of the first order electric response function $S^{(1)}(t)$, which is given by the autocorrelation of the chromophore transition dipole moment:

\begin{equation}
D_{abs}(t)\propto S^{(1)}(t)\propto \langle \mu (t) \mu (0)\rangle,
\end{equation}
where the average within the brackets is to be done over the whole system plus bath
degree of freedom. In particular, for a Hamiltonian of the form in Eq. \ref{Hamilt} and
a bath in thermal equilibrium, the expression above can be computed exactly to yield \cite{kuhn}:

\begin{equation}
S^{1}_{abs}(t)\propto e^{i\omega_{eg}t+G(t)-G(0)},
\end{equation}

where the function $G(t)$ is typically referred to as the line shape function, as it determines the shape of the absorption spectrum. Considering, for simplicity, a completely overdamped response function of the environment (corresponding to a Drude-Lorentz spectral density), we can write the time-correlation function of the transition as

\begin{figure}
\includegraphics[scale=0.45]{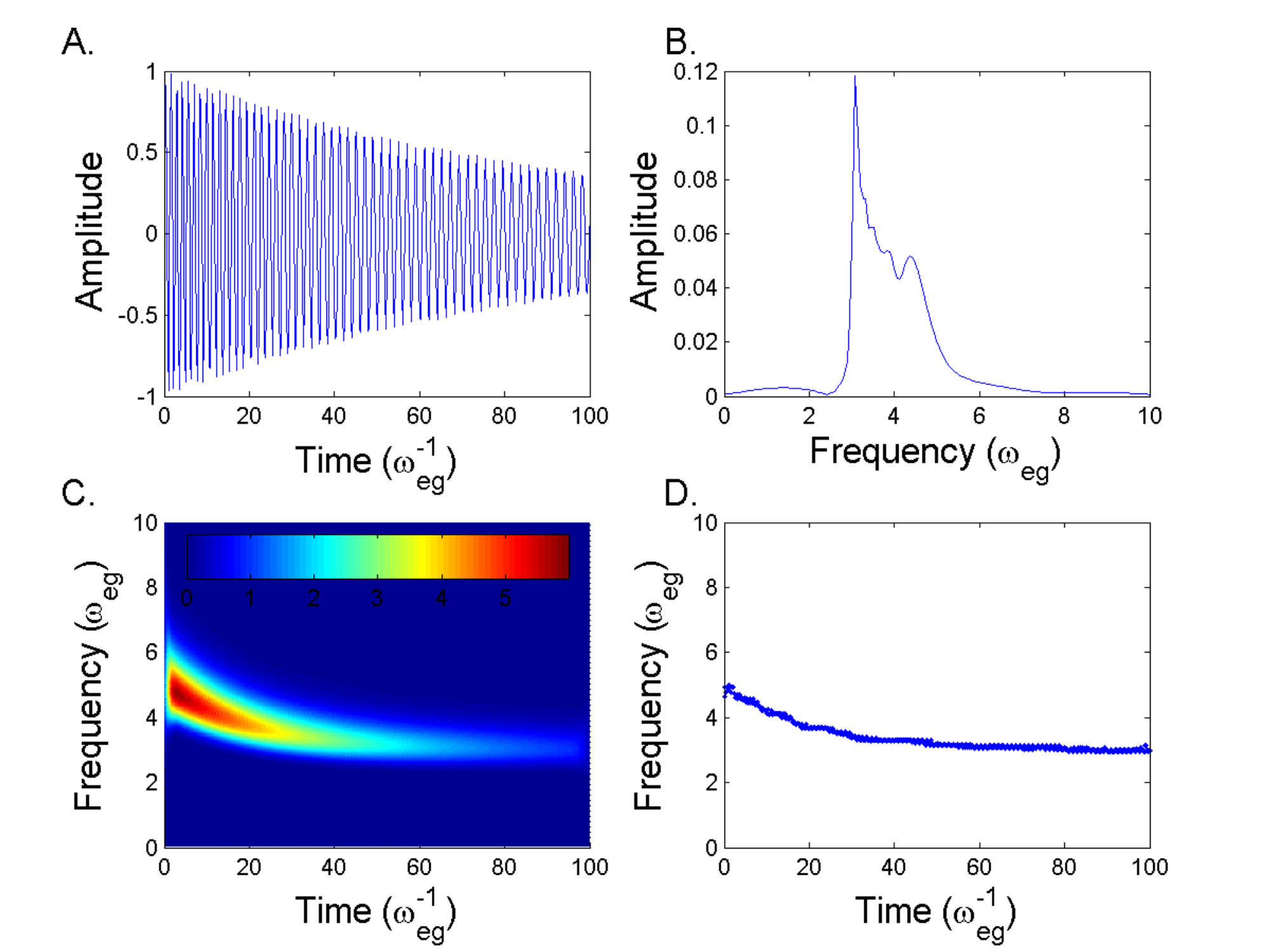}
\caption{Line Shape function for a finite damping function $G_{Re}=0.01$. Fourier Transform and wavelets analysis for the remaining parameters taking the same values as in Fig.~\ref{W0}.}
\label{W0p01}
\end{figure}

\begin{equation}
Re (S^{(1)})=e^{G_{Re} t}sin(\Omega_{eg}t+G_{Im}(t)),
\label{stokeseq}
\end{equation}
where $G(t)-G(0)=G_{Re}(t)+i G_{Im}(t)$, $G_{Im}=\lambda/\Omega_D (1-e^{-\Omega_D t})$ and $\lambda$ is the Stokes shift, or reorganization energy induced by the environment. As shown in Eq. \ref{stokeseq}, the relaxation of the molecular environment in the excited state leads to an effective change in the energy gap for vertical (optical) transitions, leading to an effective change in the peak position of the absorption spectrum with time. This occurs within a timescale set by $\tau=\Omega_{D}^{-1}$, which is the environmental relaxation time. The real part of the lineshape function ($G_{re})$ accounts for the dephasing (leading to decay) of the optical signal which is caused by fluctuations of the optical energy gaps due to random (Gausssian) environmental motion while the environment relaxes.  We now compare how these changes in the optical frequencies appear in standard FT and WT decomposition of the linear optical response of such a chromophore.

 We consider first the artificial case of zero amplitude decay ($G_{Re}=0$), whose results are shown in Fig.~\ref{W0}. In the Fourier analysis of the undamped signal in Fig.~\ref{W0}.A, we can see that the main frequency component is at the relaxed ground-excited energy gap frequency $\Omega_{ab}$. This is easily understood; the relaxation of the energy gap is complete by about $t\approx 20$, after which the undamped oscillations continue indefinitely at the relaxed frequency $\Omega_{ab}$. Higher frequency components are apparent in the FT, including a distinct second peak at a frequency $f$ of $\approx 4.5$ (Fig.~\ref{W0}.B). However, apart from the low-frequency peak at the relaxed energy, the rest of the spectrum is fairly irregular and the 'peak' at $f=4.5$ might be interpreted erroneously as another weak optical transition or a vibrational side band. The Wavelet analysis Fig.~\ref{W0}.C resolves these inconsistencies, showing a smooth evolution of the optical transition frequency from the unrelaxed energy ($f=5$) to the relaxed energy gap $(f=3)$. From this we directly infer the reorganization energy of the environment, $\lambda=2$. We note the the FT does not allow any such direct measurement of either the unrelaxed frequency or the reorganization energy. As an additional bonus, the relaxation time of the environment can be simply extracted by tracking the time evolution of the peak in the wavelet scalogram, as shown in the bottom righthand panel Fig.~\ref{W0}.D. Again, this property of the system is obscured in the FT plots.

\begin{figure}
\includegraphics[scale=0.45]{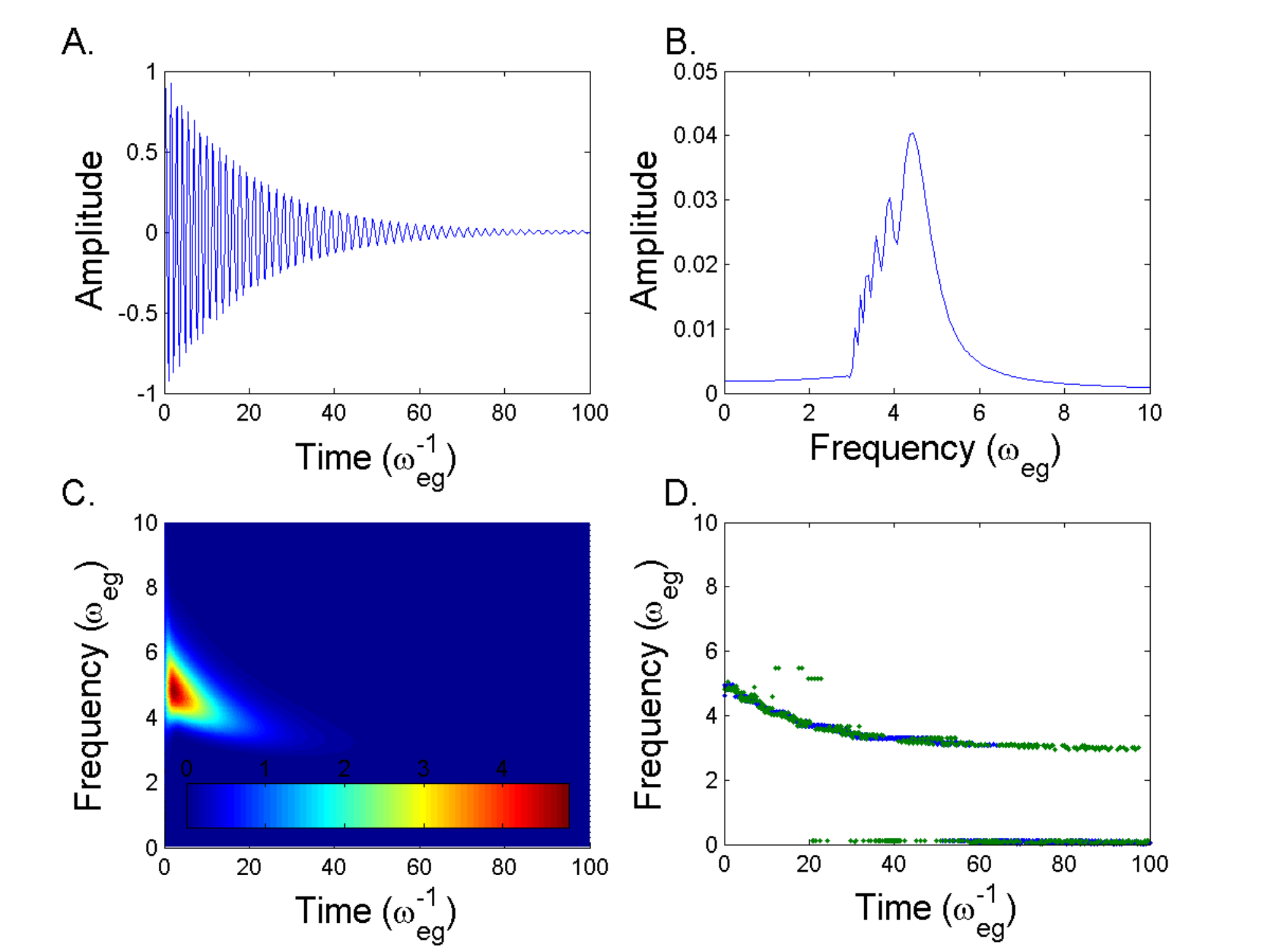}
\caption{Line Shape function for a finite damping function $G_{Re}=0.05$. Fourier Transform and wavelets analysis for the remaining parameters taking the same values as in Fig.~\ref{W0}.}
\label{W0p05}
\end{figure}

In Fig. \ref{W0p01} and  Fig. \ref{W0p05} a more realistic situation is presented, with the decay factor, $G_{Re}$ increased to values of $0.01$ and $0.05$, respectively. The signal now effectively has a finite duration, and the FT analysis consequently shows a larger contribution from the high frequency components, as they now occur over a greater fraction of the signal \cite{note}. In  Fig.~\ref{W0p01}, the lifetime of the signal is still long enough for the FT to be dominated by the relaxed frequency. However, for the enhanced damping considered in Fig. \ref{W0p05} the FT now appears as a broad peak slightly above the average value of the energy gap, with a series of sharp peaks at lower frequencies. In both these cases, the various peaks could easily be incorrectly assigned to multiple transitions or vibrational progressions.  In contrast, the wavelet scalogram and wavelet peak analysis (panels C and D) of both cases in Figs. \ref{W0p01} and \ref{W0p05} still show a smooth evolution of a \emph{single} transition, allowing the unambiguous identification of the initial unrelaxed frequency, the final relaxed frequency, the reorganization energy and the environmental relaxation time.

This example, showing how wavelets can identify time-evolving spectral lines, could have a number of other possible uses. For example, conformational changes of photoactive proteins can lead to large changes of optical transition frequencies (or even deactivate them), with the latter manifesting as 'blinking' events in absorption of fluorescence. The highly non-stationary, intermittent time series of such dynamics are naturally suited to wavelet analysis, from which conformational reorganization dynamics, and possibly their stimulation by external conditions and deliberate perturbations, might be traced. We should stress that this additional information on the time evolution of \emph{linear} optical spectra can be a powerful way to extract further information from relatively simple experiments.

\section{Conclusions}

In this work we have further highlighted the use of modern techniques from signal processing for the analysis of complex signals emanating from biological systems \cite{APP12,Sanders12} by demonstrating the usefulness of the wavelet transform analysis as a powerful technique which is well-suited to problems where the frequency spectrum is complex and
evolves in time due to the underlying physics. For dynamics resulting from multi-stage processes involving multiple states or degrees of freedom, the loss of temporal information may greatly hamper efforts to construct a kinetic or mechanistic model of the system, despite the good characterization of the individual dynamical components (excitations) that is obtained by spectral analysis. Our examples have been taken from ultrafast spectroscopy - where we believe that this technique could be especially useful - but we hope to have emphasized that this techniques can be applied to any measurement of non-stationary or transient responses in physical, chemical and biological systems. Further examples of the use of wavelets in addressing specific problems will be presented in forthcoming work.

\section{Acknowledgements}

This work was supported by an Alexander von Humboldt
Professorship, the ERC Synergy grant BioQ, the EU Integrating project SIQS and STREP PAPETS and
the Spanish Ministerio de Econom\'ia y Competitividad under Project No. FIS2012-30625. AWC acknowledges the support of the Winton Programme for the Physics of Sustainability.


\begin{thebibliography} {99}
\bibitem{note} For the undamped case, the ratio of contributions is fixed by the total data acquisition time.
\end{thebibliography}
\end{document}